\documentstyle[prl,aps,multicol,epsf]{revtex}
\begin{document}

\draft
\title{ Experimental evidence for Coulomb charging effects in an
open quantum dot at~zero~magnetic~field
}
\author{C.-T.~Liang, M.~Y.~Simmons, C.~G.~Smith, G.~H.~Kim,
D.~A.~Ritchie and M.~Pepper
}

\address{
Cavendish Laboratory,
Madingley Road,
Cambridge CB3 0HE,
United Kingdom
}

\date{\today}

\maketitle

\widetext
\begin{abstract}
\leftskip 54.8pt
\rightskip 54.8pt

We have measured the low-temperature transport properties of an open 
quantum dot formed in a clean one-dimensional channel. For the 
first time, at zero magnetic field, continuous and periodic
oscillations superimposed upon ballistic conductance steps are
observed when the conductance through the dot $G$ exceeds
$2e^2/h$. We ascribe the observed conductance 
oscillations to evidence 
for Coulomb charging effects in an open dot. This is
supported by the evolution of the oscillating features for 
$G>2e^2/h$ as a function of both temperature and barrier transparency.

\pacs{PACS numbers: 73.40.Gk, 73.20.Dx}
\end{abstract}

\begin{multicols}{2}
\narrowtext
The use of sub-micron surface gates fabricated over GaAs/AlGaAs 
heterostructures to electrostatically squeeze \cite{Trevor} the
underlying two-dimensional electron gas (2DEG) into various shapes,
has been proved a powerful tool to realize low-dimensional electron
systems. By employing this technique, it is possible to define
a quantum dot which confines electrons in an isolated region within
a 2DEG. Consider a lateral quantum dot \cite{CGSdot} weakly 
coupled to the source and drain contacts where the tunneling
conductance through the dot $G$ is low, $G\ll2e^2/h$.
If the thermal smearing $k_{B}T$ and the chemical potentials
in the leads are much smaller than the Coulomb charging energy $e^2/C$ 
which is required for adding an extra electron to the quantum dot, transport 
through the dot is inhibited. This is the Coulomb 
blockade (CB) of single electron tunneling \cite{Meirav}.  
It has been demonstrated \cite{McEuen91} that transport through small 
quantum dots is determined by Coulomb charging effects \cite{Weis} 
as well as zero-dimensional (0D)
quantum confinement effects \cite{Ash}. 

A versatile quantum dot may be defined by two pair of split-gates
which introduce two quantum point contacts acting as the entrant and
exit barriers to the dot, and two side-gates which are used to deplete
electrons within the dot \cite{Staring}. It has been reported that
at zero magnetic field Coulomb charging effects only occur when the
conductance of the two quantum point contacts, and of the quantum dot
as a whole, falls below $2e^2/h$ 
\cite{Staring,Kouwenhoven1,Williamson}. 
No oscillations were found when either or both of the quantum point
contacts were set to above $G>2e^2/h$ so that the dot was open to
the 2DEG reservoirs \cite{Staring,Kouwenhoven1,Williamson}.
Thus at present, it is widely accepted that at zero magnetic field, the 
conductance $2e^2/h$ is the upper limit for which Coulomb charging effects can 
occur\cite{Molen}. 
Nevertheless, previously we presented evidence that at 
zero magnetic field, Coulomb charging effects can occur 
even when the total conductance
through the dot is greater than $2e^2/h$, provided that the fully
transmitted one-dimensional (1D) channels only partially screens
the Coulomb charging and there is little 
inter 1D subband scattering \cite{CTL,zozu}. 
In this work the quantum dot was formed by an impurity potential, 
and precise control over the tunnel barriers was not possible, hindering 
detailed studies of this effect.

Motivated by these early results, we have designed a new type of 
open quantum dot device with tunable barriers. 
In this Letter, we report low-temperature 
conductance measurements of an open quantum dot device
in which impurity scattering is negligible. Due to the 
unique design of our devices fabricated on an ultra high-quality
high mobility electron transistor (HEMT), 
we present clear evidence, for the first time, of Coulomb
charging effects in an open quantum dot at zero magnetic field. 
This is supported by the temperature and barrier transparency
dependence of the observed periodic conductance oscillations for $G>2e^2/h$.

The two-layered Schottky gate pattern shown in the inset of Fig.~1
was defined by electron beam lithography on the surface of a high-mobility
GaAs/Al$_{0.33}$Ga$_{0.67}$As heterostructure T258, 157~nm above a 
2DEG. There is a 30~nm-thick layer of Polymethylmethacrylate (PMMA)
which has been highly dosed by an electron beam, to
act as a dielectric \cite{Zailer} between the split-gate (SG)
and three gate fingers (F1, F2, and F3) so that all gates can be 
independently controlled. The carrier concentration of the 2DEG was
$1.6\times 10^{15}$~m$^{-2}$ with a mobility of 250~m$^{2}$/Vs after 
brief illumination by a red light emitting diode. 
The corresponding transport mean free path is 16.5~$\mu$m, much longer
than the effective 1D channel length. Experiments were performed
in a dilution refrigerator at $T = 50$~mK and the two-terminal conductance 
$G=dI/dV$ was measured using an ac excitation voltage of 10~$\mu$V at a
frequency of 77~Hz with standard phase-sensitive techniques. 
In all cases, a zero-split-gate-voltage series resistance ($\approx$
900~$\Omega$) is subtracted. Two samples, at five different cooldowns, show
similar characteristics and measurements taken from one of these
are presented in this paper.

Trace 1 in figure 1 shows the conductance measurements $G(V_{SG})$ 
as a function
of split-gate voltage $V_{SG}$ when all finger gate voltages $V_{F1}$,
$V_{F2}$, and $V_{F3}$ are zero.
We observe conductance plateaus at multiples of $2e^2/h$, with no
resonant feature superimposed on top, as expected for a clean 1D
channel. When the channel is defined at
$V_{SG} = -1.132$~V, five quantized conductance steps are observed when
each one of the finger gates is swept while the others are earthed 
to the 2DEG as shown in traces 2-4 (Fig.~1). These experimental results 
demonstrate that we have a clean 1D channel in which impurity 
scattering is negligible. Periodic resonant features, as shown later, 
are only observed when large negative voltages are applied to both 
F1 and F3.

We can define a lateral quantum dot by applying voltages on 
SG, F1, and F3 while keeping F2 earthed to the 2DEG.
Trace 1 in Fig.~2 shows the gate characteristics $G(V_{SG})$ for
$V_{F1} = -1.941$~V and $V_{F3} = -1.776$~V at $T=50$~mK. Striking periodic
and continuous conductance oscillations superimposed on 
ballistic conductance steps are observed. 
We ascribe the observed conductance oscillations for $G<2e^2/h$ to Coulomb
charging effects \cite{Staring,Kouwenhoven1,Williamson}.
The observed periodic conductance oscillations for $G > 2e^2/h$
are unexpected and are the main subject of this paper.
Unlike lateral quantum dots whose tunnel barriers are
defined by two pair of split-gates, in our system, the tunnel barriers
arise from depletion from overlying finger gates. This causes a large
barrier thickness so that we do not observe well-isolated single electron
tunnelling beyond pinch-off in our case. In contrast to the well-quantized
conductance plateaus shown in Fig.~1~(a), applying voltages to F1 and F3
results in conductance steps that are not as flat or well quantized.
With the finger gates earthed to the 2DEG, the channel pinches-off at 
$V_{SG} = -1.8$~V compared with $V_{SG} = -0.7$~V when $V_{F1}= -1.941$~V
and $V_{F3}= -1.776$~V. Thus as voltages are applied to F1 and F3, the lateral
confinement weakens and the conductance steps become less pronounced.
The conductance steps also deviate from their quantized values. The most
likely reason for this effect is due to the introduction of two
tunnel barriers which enhances back-scattering in the channel, thereby
reducing the transmission probability of 1D channels \cite{Buttiker}
to be less than 1. 

Previously in a lateral quantum dot
\cite{Staring,Kouwenhoven1,Williamson} it has been 
observed that
Coulomb oscillations increase in height and decrease in width as the
conductance
decreases. This increase in height arises
from an accumulation of the electron wave function
in the dot, giving rise to resonant
coherent effects, as the dot becomes isolated
from the source and drain contacts. From figure 2 we
can see that no such increase
in height is observed in our system as the conductance is decreased.
We believe that the thicker tunnel barriers in our system make it more
difficult for the electrons
to tunnel out such that the electron lifetime within the dot becomes
so large it exceeds the inelastic scattering time. In such a situation resonant
coherent effects decrease with the result that the Coulomb oscillations
do not increase in height close to pinch-off.

We also find that the peak widths do not decrease as $G$ decreases.
Generally for $G>2e^2/h$, it is expected that the presence of a
fully transmitted
1D channel might cause mode mixing between 1D channels in the quantum dot
which smears out charging effects. However since our samples are
fabricated on an ultra high-quality HEMT it is likely that there is
little 1D mode mixing such that the level broadening for Coulomb
oscillations is similar for both cases when
$G<2e^2/h$ and $2e^2/h<G<4e^2/h$.

Having defined a quantum dot, we now calculate the dot size and
the number of electrons it contains following
the method described in Ref.~\cite{Mark}. 
For $V_{SG} = -0.5$~V, $V_{F1} = -1.941$~V, 
and  $V_{F3} = -1.776$~V, we observe Aharonov-Bohm type oscillations
as a function of applied perpendicular magnetic field
\cite{vanWees} with a period $\Delta B$ of 14.7~mT,
giving a dot area $A$ of $2.81\times 10^{-13}$~m$^{2}$.
Using the split-gate to change the dot area at a constant magnetic field
of 0.8~T, the Aharonov-Bohm period of oscillations $\Delta V_{SG}^{AB}$
was measured to be 8.772~mV \cite{Rob}. 
Thus $\Delta V_{SG}^{AB}$/$\Delta A =
1.697\times 10^{12}$~Vm$^{-2}$. Each CB oscillation corresponds to
removing an electron from the dot so that the reciprocal of the CB period
$\Delta N$/$\Delta V_{SG}^{CB}$ is $263.3$~V$^{-1}$.
From the product of these two terms we obtain the local carrier density in
the dot to be $4.468\times 10^{14}$~m$^{-2}$. Combining this value with the
dot area $A$ gives the number of electrons in the dot $N\approx$~126.
From the local Fermi energy $E_{F}^{loc}$ and the number of electrons
within the dot, we estimate the 0D confinement energy
$E_{F}^{loc}/N$ to be at most 12.4~$\mu$eV, comparable to 
the thermal smearing at 50~mK.
The reason for this is due to the large dimensions of our sample.
Therefore electron transport through our quantum dot can be described
in terms of a classical Coulomb charging picture where the 0D quantum 
confinement energy is much smaller than the Coulomb charging energy,
similar to the case of a metal.

As shown in Fig~2, for $G<2e^2/h$, the conductance oscillations persist 
up to $T=1$~K. The oscillations for $G>2e^2/h$ have a strong temperature
dependence and become indistinguishable above $T=410$~mK. 
Note that the thermal broadening $k_{B}T$ at this temperature is 
still much larger than the estimated 0D quantum confinement energy,
excluding an interpretation that conductance oscillations for $G>2e^2/h$ 
are due to tunnelling through 0D states in the quantum dot.
To determine the total capacitance between the dot and the gates of the 
sample, we measured the conductance oscillations by varying the voltage
on the different gates, while keeping the voltages on the remaining gates
fixed. From this we obtain $\Delta V_{F1} = 23.806$~mV, 
$\Delta V_{F2} = 8.683$~mV, $\Delta V_{F3} = 25.894$~mV, and
$\Delta V_{SG} = 3.593$~mV. According to this the total gate-dot
capacitance $C_{g}$ is estimated to be $7.579\times 10^{-17}$~F. 
Neglecting the capacitance between the dot and the 2DEG reservoirs, 
we calculate the Coulomb charging energy $e^2/C_{g}$ to be $0.211$~meV,
comparable to the thermal broadening at $T \approx 2$~K, which is 
consistent with the observation that close to pinch-off Coulomb
oscillations persist up to 1~K.

In order to study the unexpected presence of periodic conductance
oscillations for $G > 2e^2/h$ in more detail, we have measured 
their dependence on barrier transparency. Figure~3~(a) shows $G(V_{SG})$ 
as $V_{F1}$ and $V_{F3}$ are simultaneously decreased, thus increasing
barrier height (decreasing barrier transparency)
at zero magnetic field. Figure~3~(b) is a continuation of Fig.~3~(a) at
even more negative finger gate voltages. We number peaks in 
$G(V_{SG})$ counted from pinch-off. Note that at pinch-off, we estimate
that there are still $\approx~70$ electrons within the dot.
Consider the sixth single electron 
tunnelling peak counted from pinch-off. It is evident that as the barrier 
heights are raised by making the gate finger voltages more negative, 
the peak height decreases, and 
the peak occurs at a more positive $V_{SG}$, i.e., 
where the channel is wider, as indicated by the dashed line. 
Thus effectively we are keeping the number of electrons within
the dot constant while changing the dot shape. 
We note that the first ten tunnelling peaks counted from
pinch-off in Fig.~3~(a) gradually become indistinguishable 
as the finger gate voltages are made more negative. This is
due to the increasing barrier thickness such that
tunneling conductance becomes immeasurably small \cite{Mark}.
Over the whole measurement range, we can follow up to
48 conductance tunnelling peaks at various $V_{F1}(V_{F3})$ and
are thus able to study their barrier transparency dependence.
Note that the observed conductance
oscillations for $G>2e^2/h$ have the {\em same\/} period as that of the
oscillating features for $G<2e^2/h$.
Most importantly, as shown in Fig.~3~(a) and (b)
peaks 31-48, where $G>2e^2/h$ (shown in the uppermost
curve), all gradually evolve 
into conductance oscillations for $G<2e^2/h$ due to Coulomb 
charging \cite{Staring,Kouwenhoven1,Williamson}
as the barrier heights and thickness increase.  
This result strongly suggests that the conductance oscillations (for peak
31-48 in the uppermost curve shown in Fig.~3~(a)
) and the oscillations shown in the lowermost curves (Fig~3~(b))
are of the {\em same\/} physical origin-Coulomb charging, 
compelling experimental evidence for charging
effects in the presence of fully transmitted 1D subbands at {\em zero\/}
magnetic field. 

Finally, we ask the following question: 
Why is evidence for Coulomb oscillations for 
$G >2e^2/h$ at zero magnetic field only observed in our system?
The reason for this is unclear at present but we speculate that in our
case, the overlaying finger gates introduce abrupt tunnel barriers.
This is in contrast to work where the tunnel barriers were produced
by two quantum point contacts in which smoothly varying potentials
are introduced close to the entrant and the exit to the dot. 
The abrupt tunnel barriers in our case, might reduce the adiabaticity 
of the transmission of 1D channels, thereby enhancing electron 
back-scattering within the dot. This could cause the 
Coulomb charging effects to be more pronounced in our system as 
the tunneling 1D channels for $G>2e^2/h$ are confined within 
the open dot. Another possible reason is that 
there is very little inter-1D-subband scattering in our case
compared with other systems as endorsed by the lack of 
resonant features when the tunnel barriers are transparent.

In conclusion, we have presented low-temperature experimental
results on an open quantum dot electrostatically defined by a
split-gate, and overlaying finger gates which introduce tunnel 
barriers. Periodic and continuous oscillations 
superimposed upon ballistic conductance
steps are observed even when the conductance through the quantum dot
is greater than $2e^2/h$. At zero magnetic field, a direct
transition of conductance oscillations for $G>2e^2/h$ to those
for $G<2e^2/h$ due to Coulomb charging effects is observed with
decreasing barrier transparencies. The temperature dependence of
the observed oscillating features for $G>2e^2/h$ excludes the
interpretation that they are due to tunneling through single-particle
confinement energy states within the dot.
Both results suggest that at zero magnetic field charging effects
can occur in the presence of a fully transmitted 
1D channel, in contrast to the current experimental
and theoretical understanding of Coulomb charging.

This work was funded by the UK EPSRC, and in part, by the US Army
Research Office. We thank C.H.W. Barnes and V.I.
Talyanskii for discussions. 
G.H.K. acknowledges financial assistance from Clare College.

\centerline{\bf Figure Captions}

FIG. 1.
Trace 1: $G(V_{SG})$ for all finger gates at 0~V. 
Trace 2 to 4: $G(V_{F1})$ (in solid line), $G(V_{F3})$ (in dashed line)
and $G(V_{F2})$ (in dotted line) for $V_{SG}= -1.132$~V.
The inset shows an scanning electron micrograph of a typical
device. The brightest regions correspond to finger
gates with joining pads, labelled as F1, F2, and F3 lying above 
the split-gate (labelled as SG), 
with an insulating layer of crosslinked PMMA in between.

FIG. 2.
$G(V_{SG})$ for $V_{F1} = -1.941$~V, $V_{F2} = 0$~V, and $V_{F3} = 
-1.776$~V at various temperatures $T$.
From left to right: $T = $1, 0.5, 0.45, 0.41, 0.35, 0.3, 0.26, 0.2, 0.18, 
0.17, 0.15, 0.11, 0.09, 0.065 and $0.05$~K.
Curves are successively displaced by a horizontal offset of 0.02~V 
for clarity. 

FIG. 3.
(a) $G(V_{SG})$ at various voltages applied on F1 
and F3 at zero magnetic field. From top to bottom: 
$V_{F1} = -1.907$~V to $-1.965$~V in $2$~mV steps 
($V_{F3} = -1.733$~V to $-1.805$~V in $2.5$~mV steps) (b) 
Continuation of figure~3~(a). From 
top to bottom: $V_{F1} = -1.965$~V to $-2.023$~V in $2$~mV steps
($V_{F3} = -1.805$~V to $-1.8775$~V in $2.5$~mV steps)
Curves are successively offset by $(0.0344)(2e^{2}/h)$ 
for clarity. Conductance
tunneling peaks are numbered to serve a guide to the eye for 
the evolution of oscillating structures in $G(V_{SG})$. 
The dashed line in Fig.~3~(a) serves as a guide to the eye for the 
evolution of peak 6. The measurement temperatures were 50~mK.

\end{multicols}

\end{document}